\documentclass[twocolumn,floatfix,prl,showpacs,superscriptaddress]{revtex4}
\usepackage{graphicx}
\usepackage{amssymb}
\usepackage{amsmath}
\usepackage{color}
\usepackage{psfrag}
\usepackage{epsfig}
\usepackage{bbm}
\usepackage{bm}
\newcommand{\ket}[1]{| #1 \rangle}

\newcommand{\rb}[1]{\left( #1 \right)}

\newcommand{\ew}[1]{\langle #1 \rangle}
\newcommand{\beq}{\begin{eqnarray}}
\newcommand{\eeq}{\end{eqnarray}}

\newcommand{\op}[2]{| #1 \rangle \langle #2 |}

\newcommand{\cum}[1]{\langle\langle #1 \rangle \rangle}
\newcommand{\eq}[1]{Eq.~(\ref{#1})}

\begin{document}
\title{Frequency-dependent counting statistics in interacting nanoscale conductors}
\author{C. Emary}
\affiliation{
  Institut f\"ur Theoretische Physik,
  Hardenbergstr. 36,
  TU Berlin,
  D-10623 Berlin,
  Germany
}
\author{D. Marcos}
\author{R. Aguado}
\affiliation{
  Departamento de Teor\'ia de la Materia Condensada,
  Instituto de Ciencia de Materiales de Madrid, CSIC,
  Cantoblanco 28049, Madrid, Spain
}
\author{T. Brandes}
\affiliation{
  Institut f\"ur Theoretische Physik,
  Hardenbergstr. 36,
  TU Berlin,
  D-10623 Berlin,
  Germany
}
\date{\today}
\begin{abstract}
We present a formalism to calculate finite-frequency current
correlations in interacting nanoscopic conductors. We work within the
n-resolved density matrix approach and obtain a multi-time cumulant
generating function that provides the fluctuation statistics solely from the spectral decomposition of the Liouvillian. We apply the
method to the frequency-dependent third cumulant
of the current through a single resonant level and through a
double quantum dot. Our results, which show that deviations from
Poissonian behaviour strongly depend on frequency, demonstrate the
importance of finite-frequency higher-order cumulants in
fully characterizing transport.

\end{abstract}
 \maketitle

Following the considerable success of shot-noise in the
understanding of transport through mesoscopic systems \cite{bla00},
attention is now turning towards the higher-order statistics of
electron current.  The so-called Full Counting Statistics (FCS) of
electron transport yields all moments (or cumulants) of the
probability distribution $P(n,t)$ of the number of transferred
electrons during time $t$. Despite their difficulty, measurements of
the third moment of voltage fluctuations have been made
\cite{reu03,rez05}, and recent developments in single electron
detection \cite{fuj,gus06,Sukh07} promise to open new horizons on
the experimental side.

The theory of FCS is now well established in the zero-frequency limit
\cite{lev93,yul99,bag03}.  However, this is by no means the full picture,
since the higher-order current correlators at finite frequencies
contain much more information than their zero-frequency
counterparts. Already at second order (shot-noise), one can extract
valuable information about transport time scales and correlations.
When the conductor has various intrinsic time scales like, for
example, the charge relaxation time and the dwelling time of a chaotic
cavity \cite{nag04}, one needs to go beyond second-order in order
to fully characterize electronic transport.  Apart from this
example, and some other notable exceptions
\cite{pil04,gala03,sal06}, the behaviour of finite-frequency
correlators beyond shot-noise is still largely unexplored.


In this Rapid Communication, we develop a theory of frequency-dependent current
correlators of arbitrary order in the context of the $n$-resolved
density matrix (DM) approach, --- a Quantum Optics technique
\cite{cook81} that has recently found application in mesoscopic
transport \cite{gur98}. Within this approach, the DM of the system,
$\rho(t)$, is unravelled into components $\rho^{(n)}(t)$ in which
$n=n(t)=0,1,\ldots$ electrons have been transferred to the
collector. Considering a generic mesoscopic system with
Hamiltonian $\mathcal{H}=\mathcal{H}_S + \mathcal{H}_L +
\mathcal{H}_T$, where $\mathcal{H}_S$ and $\mathcal{H}_L$ refer to
the system and leads respectively, and provided that the Born-Markov
approximation with respect to the tunnelling term $\mathcal{H}_T$ is
fulfilled, the time-evolution of this $n$-resolved DM can be written
quite generally as
\beq
  \dot{\bm{\rho}}^{(n)}(t) = {\cal L}_0 \bm{\rho}^{(n)}(t)
  + {\cal L}_J \bm{\rho}^{(n-1)}(t)
  \label{eom}
  ,
\eeq where the vector $\bm{\rho}^{(n)}(t)$ contains the nonzero
elements of the DM, written in a suitable many-body basis.
The Liouvillian ${\cal L}_0$ describes the `continuous' evolution of
the system, whereas ${\cal L}_J$ describes the quantum jumps of the
transfer of an electron to the collector.  We make the infinite bias
voltage approximation such that the transfer is unidirectional. By
construction, this method is very powerful for studying interacting
mesoscopic systems that are weakly coupled to the reservoirs, such
as coupled quantum dots (QDs) in the Coulomb Blockade (CB) regime
\cite{gur98,bag03,aguado-brandes04} or Cooper-pair boxes \cite{CPB}.
Within this framework, our theory of frequency-dependent FCS is of
complete generality and therefore of wide applicability.

In this $n$-resolved picture, electrons are transferred to the leads via quantum
jumps and there exists no quantum coherence between states in
the system and those in the leads.  Thus, although the system itself
may be quantum, the measured current may be considered a 
classical stochastic variable and therefore amenable to classical
counting \cite{classical-counting}. This observation allows us to
derive various generalizations of classical stochastic results,
and obtain
a multi-time cumulant generating function in terms of local
propagators.
We illustrate our method by calculating the frequency-resolved third
cumulant (skewness) for two paradigms of CB mesoscopic transport:
the single resonant level (SRL) and the double quantum dot (DQD).


Equation (\ref{eom}) can be solved by Fourier transformation.
Defining $ \bm{\rho}(\chi,t) = \sum_n \bm{\rho}^{(n)}(t) e^{i n
\chi} $, we obtain $
  \dot{\bm{\rho}}(\chi,t) = M(\chi) \bm{\rho}(\chi,t)
  \label{eom2}
$, with $M(\chi)\equiv{\cal L}_0+ e^{i\chi}{\cal L}_J$.
Let $N_v$ be the dimension of $M(\chi)$, and
$\lambda_i(\chi);~i=1,\ldots,N_v$, its eigenvalues.  In the
$\chi\rightarrow 0$ limit, one of these eigenvalues,
$\lambda_1(\chi)$ say, tends to zero and the corresponding
eigenvector gives the stationary DM for the system. This single
eigenvalue is sufficient to determine the zero-frequency FCS
\cite{bag03}.  In contrast, here we need all $N_v$ eigenvalues.
Using the spectral decomposition, $ M(\chi) = V(\chi) \Lambda(\chi)
V^{-1}(\chi)$, with $\Lambda(\chi)$ the diagonal matrix of
eigenvalues and $V(\chi)$ the corresponding matrix of eigenvectors,
the DM of the system at an arbitrary time $t$ is given by \beq
  \bm{\rho}(\chi,t) = \Omega(\chi,t-t_0) \bm{\rho}(\chi,t_0),
\eeq where $
  \Omega(\chi;t)
  \equiv
  e^{M(\chi)t}
  =
  V(\chi) e^{\Lambda(\chi)t)} V^{-1}(\chi)
$ is the propagator in $\chi$-space, and $\bm{\rho}(\chi,t_0)$ is the
(normalized) state of the system at $t_0$, at which time we assume
no electrons have passed so that $\bm{\rho}^{(n)}(t_0) =
\delta_{n,0} \bm{\rho}(t_0)$ and thus $\bm{\rho}(\chi,t_0) \equiv
\bm{\rho}(t_0)$.
The propagator in $n$-space, $
  G(n,t)\equiv \int \frac{d\chi}{2\pi} e^{-in\chi} \Omega(\chi,t)
$, such that $ \bm{\rho}^{(n)}(t) = G(n,t-t_0)  \bm{\rho}(t_0) $,
for $ t>t_0$,
fulfils the property:
$  G(n-n_0,t-t_0)=
   \sum_{n'}G(n-n',t-t')
G (n'-n_0,t'-t_0)$,
for $t>t'>t_0$, ($n'\equiv n(t')$, $n_0\equiv n(t_0)$). This is
an operator version of the Chapman-Kolmogorov equation
\cite{stochast}.

The joint probability of obtaining $n_1$ electrons after $t_1$ and
$n_2$ electrons after $t_2$, namely $P^>(n_1,t_1;n_2,t_2)$ (the
superscript `$>$' implies $t_2 > t_1$), can be written in terms of
these propagators by evolving the local probabilities
$P(n,t)=\mathrm{Tr}\lbrace \bm{\rho}^{(n)}(t) \rbrace$ \cite{fn1}, and taking
into account $P(n_2,t_2)=\sum_{n_1} P^>(n_1,t_1;n_2,t_2)$,
such that
\beq
  P^>(n_1,t_1;n_2,t_2) &=& \mbox{\rm Tr}
  \left\{
    G(n_2-n_1,t_2-t_1)
  \right.
  \nonumber \\
  &&~~~~~~~
  \left.
  \times G (n_1,t_1-t_0)\bm{\rho}(t_0)
  \right\}
  \label{P>2}
  .
\eeq
The total joint probability reads $P(n_1,t_1;n_2,t_2) =
\mathcal{T}P^>(n_1,t_1;n_2,t_2) = P^>(n_1,t_1;n_2,t_2)
\theta(t_2-t_1) + P^<(n_1,t_1;n_2,t_2) \theta(t_1-t_2)$ where $\mathcal{T}$ is the time-ordering operator, $\theta(t)$ the
unitstep function defined as $\theta(t)=1$ for $t\ge0$, and zero otherwise, and where $P^<(n_1,t_1;n_2,t_2)$ is the joint probability with $t_2 < t_1$. It should be noted that, in contrast to the
local probability $P(n,t)$, the joint probability
$P(n_1,t_1;n_2,t_2)$ contains information about the correlations at \textit{different times}.

Result (\ref{P>2}) may be alternatively derived using the Bayes
formula for the conditional density operator \cite{korotkov}:
\beq
  &&
  \!\!\!\!\!\!\!\!\!\!\!\!
  P^>(n_1,t_1;n_2,t_2)
  =
  P(n_1,t_1)P^>(n_2,t_2\vert n_1,t_1)
  \nonumber \\
  &=& \mathrm{Tr}\Big\lbrace \bm{\rho}^{(n_1)}(t_1)
  \Big\rbrace \mathrm{Tr}\Big\lbrace \bm{\rho}^{(n_2\vert n_1)}(t_2)\Big\rbrace
  \nonumber \\
  &=& \mathrm{Tr}\Big\lbrace \bm{\rho}^{(n_1)}(t_1) \Big\rbrace \mathrm{Tr}\Bigg\lbrace
  G(n_2-n_1,t_2-t_1)\frac{\bm{\rho}^{(n_1)}(t_1)}{\mathrm{Tr}\Big\lbrace
  \bm{\rho}^{(n_1)}(t_1)\Big\rbrace}\Bigg\rbrace.
  \nonumber
\eeq The normalization in the denominator accounts for the collapse
$n=n_1$ at $t=t_1$ using Von Neumann's projection postulate
\cite{korotkov}. Equation (\ref{P>2}) is recovered when
$\bm{\rho}^{(n_1)}(t_1)$ is written as a time evolution from $t_0$.

The two-time cumulant generating function (CGF) associated with
these joint probabilities is \beq
e^{-\mathcal{F}(\chi_1,\chi_2;t_1,t_2)}
  &\equiv& \sum_{n_1,n_2}
  P(n_1,t_1;n_2,t_2)
  e^{i n_1 \chi_1 + i n_2 \chi_2},\nonumber
\eeq which, using \eq{P>2}, and $e^{-\mathcal{F}}=e^{-{\cal
T}\mathcal{F}^>}={\cal T} e^{-\mathcal{F}^>}$, gives
\beq
  e^{-{\cal F}(\chi_2,\chi_1;t_2,t_1)}
  &=& {\cal T}~
  \mbox{\rm Tr}
  \left(
    \Omega(\chi_2,t_2-t_1)
  \right.
  \nonumber \\
  &&~~
  \left.
  \times\Omega(\chi_1+\chi_2,t_1-t_0)\bm{\rho}(t_0)
  \right)
  .
\eeq
The above procedure can be easily generalized to obtain the N-time
CGF, which reads:
\beq\label{genF}
  e^{-\mathcal{F}(\bm{\chi};\bm{t})}
  &=&
  \mathcal{T}~\mathrm{Tr}
  \left\{
  \prod_{k=1}^N
    \Omega
    (\sigma_k; \tau_{N-k})
    \bm{\rho}(t_0)
  \right\},
\eeq
where $\sigma_k \equiv \sum_{i=N+1-k}^N \chi_i$,
$\bm{\chi}\equiv(\chi_1,\dots,\chi_N)$, $\bm{t}\equiv(t_1,\dots,t_n)$ and
$\tau_k \equiv t_{k+1}-t_k$ \cite{biyect}. The multi-time CGF in \eq{genF} contains
a product of local-time propagators, and expresses the Markovian character of the
problem. It allows one to obtain all the
frequency-dependent cumulants from the spectral decomposition of
$M(\chi$).
The $N$-time current-cumulant ($e= 1$) is calculated using:
\beq
  &&S^{(N)}(t_1,\dots,t_N)
  \equiv \langle \delta I(t_1)\dots \delta I(t_N) \rangle = \nonumber \\
  &&=
  \partial_{t_1}\dots
  \partial_{t_N} \cum{n(t_1)\dots n(t_N)} = \nonumber \\ &&= -(-i)^N
  \partial_{t_1}\dots \partial_{t_N} \partial_{\chi_1}\dots
  \partial_{\chi_N}
  \mathcal{F}(\bm{\chi};\bm{t})\Big|_{\bm{\chi}=\bm{0}}.
  \label{corr}
\eeq
The Fourier transform of $S^{(N)}$ with respect to the time
intervals $\tau_k$, gives the $N$th-order correlation functions as
functions of $N-1$ frequencies. In particular, the
frequency-dependent skewness is a function of two frequencies which,
as a consequence of time-symmetrization and the Markovian
approximation, has the symmetries $S^{(3)}(\omega,\omega') =
S^{(3)}(\omega',\omega) =
S^{(3)}(\omega,\omega-\omega')=S^{(3)}(\omega'-\omega,\omega')=
S^{(3)}(-\omega,-\omega')$, and is therefore real.  The
$N$th-order Fano-factor is defined as $F^{(N)} \equiv S^{(N)}/\ew{I}$.

In the case where the jump matrix ${\cal L}_J$ contains a single
element, $({\cal L}_J)_{ij} =
\Gamma_R\delta_{i\alpha}\delta_{j\beta}$, which is the case for a
wide class of models including our two examples below, all the
correlation functions can be expressed solely in terms of the
eigenvalues $\lambda_k$ of $M(0)$, and the $N_v$ coefficients $
  c_k \equiv (V^{-1}{\cal L}_J V)_{kk} = \Gamma_R V_{\beta k }V^{-1}_{k \alpha}
$. The second-order Fano factor then has the simple, general form
\beq
  F^{(2)}(\omega)
 & =& 1 - 2\sum_{k=2}^{N_v} \frac{c_k
  \lambda_k}{\omega^2+\lambda_k^2},
 \label{F2}
\eeq
which has also been derived in other ways \cite{her93fli05}.
The skewness has the form $
  F^{(3)}(\omega,\omega') =
  -2
  +\sum_{i=1}^3 F^{(2)}(\nu_i)+
  \widetilde{F}^{(3)}(\omega,\omega')
$, with  $\nu_1 = \omega$, $\nu_2 = \omega'$, and $\nu_3 =
\omega-\omega'$. $\widetilde{F}^{(3)}(\omega,\omega')$ is an
irreducible contribution, the form of which is too cumbersome to be given
here. The high-frequency limit of the skewness is
$F^{(3)}(\omega,\infty) = F^{(2)}(\omega)$.

As a first example we consider a SRL, described by $\bm{\rho}=\left(
\rho_{00},\rho_{11} \right)^T$ and
$  M(\chi) =
  \rb{
  \begin{array}{cc}
    - \Gamma_L & e^{i\chi}\Gamma_R \\
    \Gamma_L & -\Gamma_R
  \end{array}
  }
  $,
in the basis of `empty' and `populated' states, $\left\lbrace\ket{0}, \ket{1}\right\rbrace$.
Employing \eq{genF}, we obtain the known results for the current
and noise, and arrive at our result for the skewness:
\beq
  F^{(3)}(\omega,\omega')
  &=&
   1
   -2\Gamma_L \Gamma_R
   \frac{
     \prod_{i=1}^2
     \rb{\gamma_i + \omega^2 - \omega \omega' + \omega'^2}
   }
   {
     \prod_{j=1}^3
     \rb{\Gamma^2+\nu_j^2}
   },
\nonumber 
\eeq
%
\begin{figure}[t]
  \begin{center}
  \epsfig{file=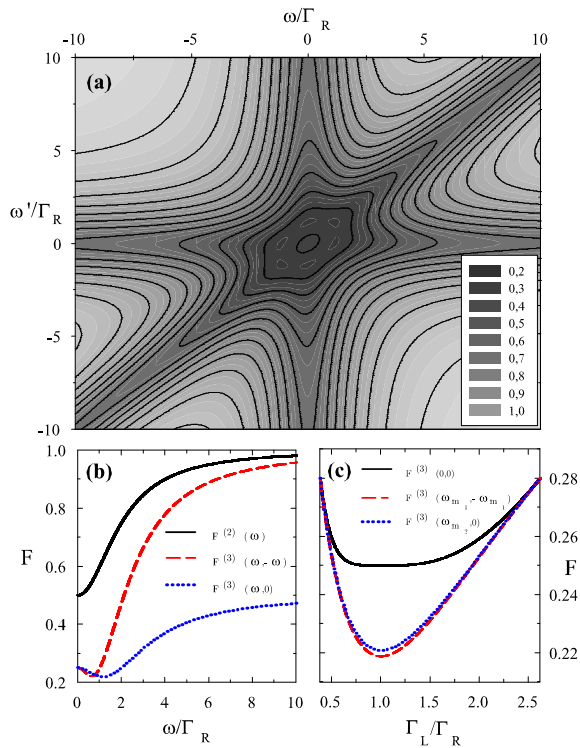, clip=true,width=0.88\linewidth}
  \caption{
     The third-order frequency-dependent Fano factor $F^{(3)}(\omega,\omega')$ for the single resonant level.
     {\bf (a)} Contour plot of $F^{(3)}(\omega,\omega')$ as function of its arguments for $\Gamma_R=\Gamma_L$.
    {\bf (b)}  Sections $F^{(3)}(\omega,0)$ and $F^{(3)}(\omega,-\omega)$
    show that the skewness is suppressed throughout frequency space both with respect to the Poissonian value of unity and to the noise
    $F^{(2)}(\omega)$. In contrast to the shotnoise, the skewness has a minimum at a finite frequency $\omega_m$, which exists in the coupling range
    $\left( 3-\sqrt{5} \right)/2 \leq \Gamma_L/\Gamma_R \leq \left( 3+\sqrt{5} \right)/2$.
    In direction $\omega=-\omega'$, this
minimum occurs, for $\Gamma_L=\Gamma_R$, at $\omega_{m1}/\Gamma_R=\sqrt{\sqrt{10}-2}/\sqrt{3}$
whereas for $\omega'=0$ the position of the minimum
shifts slightly  to $\omega_{m2}/\Gamma_R=2/\sqrt{3}$.
    {\bf (c)} The maximum suppression occurs at $\Gamma_L=\Gamma_R$.
    \label{SLSfig}
 }
  \end{center}
\end{figure}
with $\gamma_1 =\Gamma_L^2 + \Gamma_R^2$, $\gamma_2
=3\Gamma^2$, and $\Gamma=\Gamma_L+\Gamma_R$.  The zero-frequency limit $F^{(3)}(0,0)$ agrees with Ref.~\cite{jong96}.

This skewness is plotted in Fig.~\ref{SLSfig}, from which the
six-fold symmetry of $F^{(3)}$ is readily apparent.
The third-order Fano factor gives, in accordance with the noise,
a sub-Poissonian behaviour for all frequencies.
This can be easily understood as a CB suppression of the long tail of the
probability distribution of instantaneous current: due to the infinite bias,
the distribution is bounded on the left by zero, but, in principle, is not bounded on
the right (large, positive skewness). CB suppresses large current
fluctuations which explains a sub-Poissonian skewness.

Along the symmetry lines in frequency space with
$\omega'=0$, $\omega'=\omega$ and $\omega=0$, the skewness is highly
suppressed. In contrast with the noise, the minimum in the skewness
occurs at finite frequency (Fig.~\ref{SLSfig}b) with the strongest
suppression occuring at $\Gamma_L=\Gamma_R$ (Fig.~\ref{SLSfig}c).

As a second example we consider a DQD in the strong CB regime
\cite{sto96,Brandes04}. In the basis of `left' and `right' states
$\ket{L}$ and $\ket{R}$, which denote states with one excess
electron with respect to the many body `empty' state $\ket{0}$, the
Hamiltonian reads ${\cal H}_\mathrm{S} = \epsilon
\rb{\op{L}{L}-\op{R}{R}} + T_c\rb{\op{L}{R}+\op{R}{L}} $, with
detuning $\epsilon$ and coupling strength $T_c$. The two levels
$\ket{L}$, $\ket{R}$, are coupled to their respective leads with
rates $\Gamma_L$ and $\Gamma_R$.
The DM vector is now $\bm{\rho}=\left(\rho_{00},
\rho_{LL},\rho_{RR}, \mathrm{Re}(\rho_{LR}),
\mathrm{Im}(\rho_{LR})\right)^T$, and the Liouvillian in this basis
reads: \beq
  M(\chi) =
  \rb{
  \begin{array}{ccccc}
    -\Gamma_L & 0 & e^{i\chi} \Gamma_R & 0 &0\\
    \Gamma_L & 0 & 0 & 0 & - 2 T_c \\
    0 & 0 & -\Gamma_R & 0 & 2 T_c \\
    0 & 0 & 0 & -\frac{1}{2}\Gamma_R & 2 \epsilon\\
    0 & T_c & - T_c &  -2 \epsilon &  - \frac{1}{2}\Gamma_R
  \end{array}
  }
  . \nonumber
\eeq

\begin{figure}[t]
  \begin{center}
  \epsfig{file=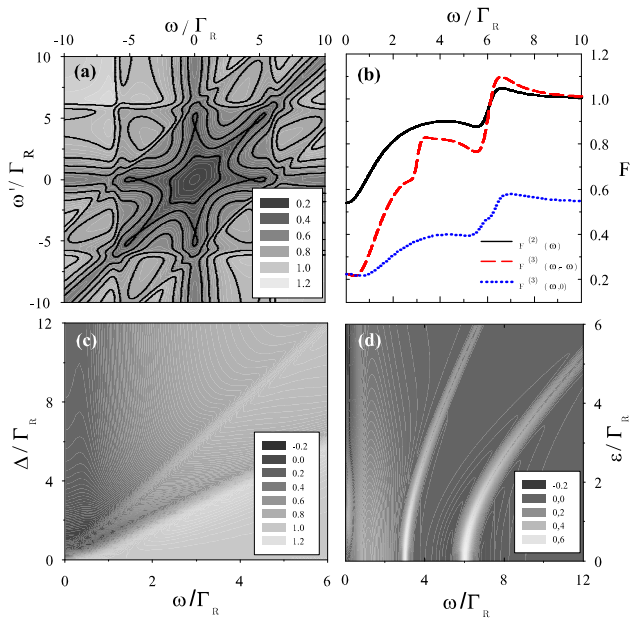, clip=true,width=0.98\linewidth}
  \caption{
    Frequency-dependent Fano skewness for the double quantum dot in Coulomb blockade.
    {\bf (a)} Contour plot in the strong coupling regime, $T_c=3\Gamma_R$,  with $\Gamma_L=\Gamma_R$ and $\epsilon=0$.
    {\bf (b)}  Sections $F^{(3)}(\omega,0)$ and $F^{(3)}(\omega,-\omega)$, and
    shotnoise $F^{(2)}(\omega)$ show a series of abrupt increases with increasing $\omega$.
    Both noise and skewness exhibit both sub- and super- Poissonian behaviour
   {\bf (c)} Varying the internal coupling $T_c$, the skewness shows rapid increases along the lines $\omega=\Delta$ and $\omega=\Delta/2$.
   For $\omega >\Delta$ the system is Poissonian (slightly super-Poissonian for $\omega \gtrsim \Delta$), while for $\omega<\Delta$ the transport is always
   sub-Poissonian. The skewness is strongly suppressed at low frequencies.
   {\bf (d)} The derivative $dF^{(3)}(\omega,-\omega)/d\omega$ as a function of frequency and detuning $\epsilon$ for $T_c=3\Gamma_L=3\Gamma_R$. Resonances occur at $\omega=\Delta$, $\Delta/2$ and $\sim\Gamma_R$.
    \label{DQDfig}
    }
  \end{center}
\end{figure}

Comparison of the quantum-mechanical level-splitting $\Delta
\equiv 2\sqrt{T_c^2 + \epsilon^2}$ with the incoherent rates
$\Gamma_{L,R}$ divides the dynamical behaviour of the system into
two regimes.
For $\Delta \ll \Gamma_{L,R}$, all eigenvalues of $M(0)$ are real, 
and correspondingly, the noise and skewness are
slowly-varying functions of their frequency arguments.
In this regime, dephasing induced by the leads suppresses
interdot coherence, and the transport is largely incoherent. In the
opposite regime, $\Delta \gg \Gamma_{L,R}$, two of the eigenvalues
form a complex pair, $ \lambda_\pm \approx \pm i \Delta -
\Gamma_R/2 +O(\Gamma/\Delta)^2$,
which signals the persistence of coherent oscillations in the
dots.  The finite-frequency correlators then show resonant
features at 
$\Delta$, since these eigenvalues enter into the denominators, as in
\eq{F2}, giving rise
to poles such as $\omega \mp \Delta - i\Gamma_R/2$.
The structure of the skewness is similar to the SRL for
weak coupling ($\Delta \ll \Gamma_{L,R}$), but much richer in the
strong coupling regime ($\Delta \gg \Gamma_{L,R}$) (Fig.~\ref{DQDfig}). 
Now the skewness
exhibits a series of rapid increases. From the origin
outwards in the $\omega$-$\omega'$ plane, we observe first a
minimum at finite frequency and then inflexion points at
$\omega\sim\Gamma_R$, $|\omega|=\Delta$, $|\omega'|=\Delta$ and
$|\omega-\omega'|=\Delta$.  Fig.~\ref{DQDfig}b shows 
sections in the $\omega$-$\omega'$ plane, and the resonant behaviour, 
in the form of Fano shapes, is most pronounced in $F^{(3)}(\omega,-\omega)$.
Starting from high frequencies, the onset of antibunching
occurs at $\omega=\Delta$. At higher frequencies, the
system has no information about correlations and is Poissonian.
The overall behaviour is seen in Fig.~\ref{DQDfig}(c) where
we plot $F^{(3)}(\omega,-\omega)$ as a function of $T_c$ and
$\omega$. The line $\omega=\Delta$ delimits two regions: at
high frequencies the skewness is Poissonian. At resonance, and after
a small super-Poissonian region at $\omega \gtrsim \Delta$, the
system becomes sub-Poissonian (and even negative, for certain internal couplings).
In the limit $\omega\rightarrow 0$, our results qualitatively agree
with those of Ref. \cite{Kiess06} for a noninteracting DQD: as a
function of $T_c$, the skewness presents two minima and a maximum
(where the noise is minimum, not shown). In our case, however, the
maximum occurs around $\Delta=\Gamma_R/2$ --- half that of the noninteracting case. 
Finally, we plot
$dF^{(3)}(\omega,-\omega)/d\omega$ as a function of both $\epsilon$
and $\omega$ (Fig.~\ref{DQDfig}d) where the resonances at
$\omega=\Delta$, $\omega=\Delta/2$ and $\omega\sim\Gamma_R$ are
clearly resolved. In contrast, the
derivative $dF^{(3)}(\omega,0)/d\omega$ (not shown), exhibits a
minimum at $w=\Delta$ for small $\epsilon$, which transforms into a
maximum for $\epsilon \gg T_c$. As expected, transport tends to be more
Poissonian as $\epsilon$ increases, signaling lost of coherence.

Despite the simplicity of the models we have studied, our results
demonstrate the importance of finite-frequency studies. Deviations
from Poissonian behaviour of higher-order cumulants are
frequency-dependent, such that a comprehensive analysis in the
frequency domain is needed in order to fully characterize
correlations and statistics in electronic transport.

Work supported by the WE Heraeus foundation, DFG grant BR 1528/5-1, and by the grants
MAT2006-03741 and FPU AP2005-0720 (MEC-Spain), 20060I003
(CSIC) and 200650M047 (CAM).



\begin{thebibliography}{99}
\bibitem{classical-counting}This is in contrast with quantum
detection, where the non-commutativity of current at different times must be explicitly
considered \cite{lev93}.
\end{thebibliography}
\end{document}